
\documentstyle{amsppt}
%
%
\catcode`\@=11
\def\logo@{}
\catcode`\@=13
%

\def\gr#1{{\goth #1}}


	
	\def\grb{{\gr b}}
	\def\grc{{\gr c}}
	\def\grd{{\gr d}}

	\def\grg{{\gr g}}

	\def\grk{{\gr k}}
	\def\grl{{\gr l}}
	\def\grm{{\gr m}}
	
	\def\gro{{\gr o}}
	\def\grp{{\gr p}}

	\def\grs{{\gr s}}
	
	\def\gru{{\gr u}}


\def\sig{\sigma}

\def\nchi{\hbox{\raise 2.5pt\hbox{$\chi$}}}


\def\CalD{{\Cal D}}

\def\CalF{{\Cal F}}

\def\CalL{{\Cal L}}

\def\CalN{{\Cal N}}


		\def\bfR{{\bold R}}


\def\ctil{\tilde c}

\def\ltil{\tilde l}

\def\qtil{\tilde q}

\def\Ftil{\widetilde F}

\def\Jtil{\widetilde J}

\def\Wtil{\widetilde W}


		\def\Fbar{\overline{F}}
		\def\Gbar{\overline{G}}
\def\hbar{\bar h}

\def\pbar{\bar p}

\def\ubar{\bar u}		
		
\def\wbar{\bar w}		
		\def\Xbar{\overline{X}}
		\def\Ybar{\overline{Y}}


		\def\Xhat{\widehat {\mathstrut X}}
		\def\Yhat{\widehat {\mathstrut Y}}

\def\rarrow{\rightarrow}

%
%

\def\r1{\sqrt{-1}}
\def\lb{\linebreak}

\def\lbd{\lambda}
\def\mx{\pmatrix}
\def\emx{\endpmatrix}
\def\smx{\left(\smallmatrix}
\def\esmx{\endsmallmatrix\right)}
\def\cx{\text{\bf C}}
\def\rx{\bfR}

\def\sst{\scriptstyle}
\def\dst{\displaystyle}
\def\sn{ \sum_{i=1}^n}
\def\tp{^{\text{\bf T}}}
\define\union#1#2{\overset{#2}\to{\underset{#1}\to \cup}}
\def\lo{\Cal L_{0}}
\def\li{\Cal L_{1}}
\def\lii{\Cal L_{2}}
\def\Ln{\Cal L}
\def\iul{I_{1,l}}
\def\ce{^{l_i c_i}}

\def\ak{\alpha_k}
\def\be{\beta_i}
\def\da{\delta_i}
\def\bj{\beta_j}
\def\dj{\delta_j}
\def\rj{\rho_j}
\def\gki{\gamma_{\kappa_i}}
\def\gkti{\gamma_{\kappa_{2i}}}
\def\gkj{\gamma_{\kappa_j}}

\def\dbi{^{\da\be}_i}
\def\Mk{{\Cal M}^{\bold k}}
\def\xc{\hat X_i}
\def\xv{\Check X_i}
\def\yc{\hat Y_i}
\def\yv{\Check Y_i}

\def\xjb{\Xbar_j}

\def\xdb{\Xbar_{2i}}

\def\ydb{\Ybar_{2i}}

\def\wu{w_{2i-1}}

\def\wz{w_{2i}}
\def\wzb{\overline{w}_{2i}}
\def\wjb{\overline{w}_j}
\def\wt{\wtil}

\def\xd{X_{2i}}
\def\xu{X_{2i-1}}
\def\yd{Y_{2i}}
\def\yu{Y_{2i-1}}
\def\ythb{\overline{\Yhat}_{2i}}
\def\yth{\Yhat_{2i}}
\def\xdc{\hat X_{2i}}
\def\xdcb{\overline{\hat X}_{2i}}
\def\xth{\Xhat_{2i}}
\def\xthb{\overline{\Xhat}_{2i}}
\def\xuc{\hat X_{2i-1}}
\def\xucb{\overline{\hat X}_{2i-1}}

\def\xjc{\hat X_j}
\def\xjcb{\overline{\hat X}_j}
\def\xjhb{\overline{\Xhat}_j}
\def\xdv{\Check X_{2i}}
\def\xdvb{\overline{\Check X}_{2i}}
\def\xuv{\Check X_{2i-1}}
\def\xuvb{\overline{\Check X}_{2i-1}}
\def\xjv{\Check X_j}
\def\xjvb{\overline{\Check X}_j}
\def\sj{ \sum_{j=1}^n}
\def\sm{ \sum_{i=1}^m}
\def\sjm{ \sum_{j=2m+1}^n}

\def\qt{\qtil}
\def\pt{\tilde p}
\def\vPhi{\varPhi}
\def\tr{\text{tr}}
\def\al{\alpha_i}
\def\ai{\al}
\def\adi{\alpha_{2i}}
\def\adiu{\alpha_{2i-1}}
\def\aj{\alpha_j}
\def\adib{\overline{\alpha}_{2i}}
\def\adiub{\overline{\alpha}_{2i-1}}
\def\ajb{\overline{\alpha}_j}
\def\kmm{\kern -6pt}
\def\ktm{\kern 7pt}
\def\kmc{\kern -7pt}
\def\ktmi{\kern 5pt}
\def\kmmi{\kern -6pt}
\def\kenmi{\kern 3pt}
\def\knenmi{\kern -4pt}

\def\albar{\overline\alpha}

\def\vs{\vspace{4mm}}

\redefine\cite#1{{\bf[#1]}}

\def\Ad{\text{\rm Ad}}

\def\contil#1{\kern2pt\bar{\kern-2pt{\tilde #1}}}
\def\ncaliu{\CalN^{(1)}_i}
\def\ncalid{\CalN^{(2)}_i}
\def\ncalj{\CalN_j}
\def\ncali{\CalN_i}

\def \smaller {\eightpoint}
\def \s {\sigma}
\def\tr{\operatorname{tr}} 
\def\pr{\operatorname{pr}} 
%
%
\TagsOnRight
\parindent=8 mm
\magnification \magstep1
\hsize = 6.25 true in
\vsize = 8.5 true in
\hoffset = .2 true in
\parskip=\medskipamount
%
%
\def\mdbox#1{\hbox{$\dst #1$}}
\def\msbox#1{\hbox{$\sst #1$}}

\def\contil#1{\kern2pt\bar{\kern-2pt{\tilde #1}}}
\def\espho{height3pt&\omit&&\omit&&\omit&&\omit&&\omit&&\omit&&\omit&\cr}

\def\alargeline{\espho\noalign{\hrule height 0.8pt}\espho}

\def\boxedalign{\bgroup\offinterlineskip\vbox\bgroup
\hrule height 1.2pt\tabskip=0pt
\halign to \hsize
\bgroup &\vrule width 1.2pt##
\tabskip=3pt plus 10pt minus 1pt&
\hfil$\vcenter{##}$\hfil&\vrule width 0.8pt##&
\hfil$\vcenter{##}$\hfil&\vrule width 0.8pt##&
\hfil$\vcenter{##}$\hfil&\vrule width 0.8pt##&
\hfil$\vcenter{##}$\hfil&\vrule width 0.8pt##&
\hfil$\vcenter{##}$\hfil&\vrule width 0.8pt##&
\hfil$\vcenter{##}$\hfil&\vrule width 0.8pt##&
\hfil$\vcenter{##}$\hfil&\tabskip=0pt\vrule
width 1.2pt##\cr }
\def\endboxedalign{\cr\espho\egroup\hrule height
1.2pt\egroup\egroup }

%
%
%
\topmatter
\title
 Matrix Nonlinear Schr\"odinger Equations \\
and Moment Maps into Loop Algebras${}^\dag$
\endtitle
\rightheadtext{Darboux Coordinates and CNLS Equations}
\leftheadtext{J. Harnad and M.-A.~Wisse}
\author
J. Harnad${}^1$
and
M.-A. Wisse${}^2$
\endauthor
\endtopmatter
\footnote""{${}^1$Department of Mathematics and
  Statistics, Concordia University, Montr\'eal, P.Q. and \lb
  Centre de Recherches Math\'ematiques,
  Universit\'e de Montr\'eal, C.P. 6128-A,
  Montr\'eal, P.Q. , Canada H3C 3J7}
\footnote"" {${}^2$D\'epartement de math\'ematiques
et de statistique, Universit\'e de Montr\'eal, C.P. 6128-A,\lb
  Montr\'eal, P.Q., Canada H3C 3J7}
\footnote"" {${}^\dag$Research supported in part by the
Natural Sciences Engineering Research Council of Canada and the Fonds FCAR
du Qu\'ebec}
\bigskip
\centerline{\bf Abstract}
\bigskip
\baselineskip=10pt
\centerline{
\vbox{
\hsize= 5.5 truein
{\smaller It is shown how Darboux coordinates on a reduced symplectic vector
space may be used to parametrize the phase space on which the finite
gap solutions of matrix nonlinear Schr\"odinger equations are realized as
isospectral Hamiltonian flows. The parametrization follows from a moment map
embedding of the symplectic vector space, reduced by suitable group actions,
into the dual $\tilde\grg^{+*}$ of the algebra $\tilde\grg^+$ of positive
frequency loops in a
Lie algebra $\grg$. The resulting phase space is identified with a Poisson
subspace of $\tilde\grg^{+*}$ consisting of elements that are rational in the
loop parameter. Reduced coordinates associated to the various Hermitian
symmetric Lie algebras $(\grg,\grk)$ corresponding to the classical Lie
algebras are obtained.
}}}
\baselineskip=16pt \bigskip \bigskip

\def\lra{\longrightarrow}
\def \wt{\widetilde}
\document
\bigskip
\noindent {\bf 1. Introduction}
\medskip

In a series of recent papers \cite{AHP, AHH1, AHH2} a systematic method was
developed for parametrizing quasi-periodic solutions to integrable systems
of PDE's in terms of Darboux coordinates on a finite dimensional phase space.
The general approach consists of reducing symplectic vector spaces under
suitable continuous or discrete Hamiltonian group actions, and embedding the
reduced spaces as Poisson submanifolds in the dual $\grg^{+*}$ of the positive
frequency part of a loop algebra $\tilde\grg$. The image space consists of
loops
in $\grg$ extending as rational functions of the loop parameter, and the
PDE's in question arise as the compatibility conditions for  pairs of
commutative isospectral Hamiltonian flows induced by spectral invariants.
A number of examples have been studied within this framework, including the
cubically nonlinear Schr\"odinger (NLS) equation and the coupled, two-component
nonlinear Schr\"odinger (CNLS) equation.

In \cite{FK}, Fordy and Kulish  gave a list of
generalized nonlinear Schr\"odinger equations
related to the Hermitian symmetric spaces correponding to
classical Lie algebras, together with their associated matrix
Lax pair commutative flows.  The general form
for such equations is
$$
\align
\r1 q_t &= q_{xx} + 2qpq  \\
-\r1 p_t &= p_{xx} + 2pqp\, ,  \tag{1.1}
\endalign
$$
where $q$ and $p\tp$ are complex $a\times b$ matrices.
Specific cases are obtained by imposing further invariance
conditions under involutive automorphisms.

In Hamiltonian terms, the
corresponding Lax equations may be viewed as repesenting flows in
the dual of the
positive part $\wt{\grs \grl}(r,\cx)^+$ of the loop algebra
$\wt{\grs \grl}(r,\cx)$, or subalgebras obtained as fixed point sets
under involutive automorphisms.
Using the  Lie Poisson bracket structure
on  $\wt{\grs \grl}(r,\cx)^{+*}$, the Adler-Kostant-Symes (AKS)
theorem gives the Lax pair representations:
$$
\align
\frac d{dx}\Cal L(\lbd) &=
[(\lbd \Ln (\lbd))_+,\Cal L(\lbd)] \tag{1.2a} \\
\frac d{dt}\Cal L(\lbd) &=
[(\lbd^2 \Ln (\lbd))_+,\Cal L(\lbd)] \tag{1.2b}
\endalign
$$
for Hamilton's equations corresponding to suitably chosen elements
of the  Poisson commuting ring
$\CalF \equiv I(\wt{\grs \grl}(r,\cx)^{*})|_{\wt{\grs
\grl}(r,\cx)^{+*}}$ of $Ad^*$-invariant functions on $\wt{\grs
\grl}(r,\cx)^*$, restricted to the subspace $\wt{\grs
\grl}(r,\cx)^{+*}$. Here $\lbd$ is the loop parameter and $\CalL(\lbd)$ is
chosen as holomorphic outside a disc $D$ in the complex $\lbd$-plane centered
at $0$. The $+$ subscript in \thetag{1.2a,b} means projection to the
positive part $\wt{\grs \grl}(r,\cx)^+$ of the loop algebra (i.e.
loops that extend holomorphically to the interior of the circle
$S^1=\partial D$).
As usual, there is an identification understood between  $\wt{\grs
\grl}(r,\cx)$ and a dense subspace of  $\wt{\grs \grl}(r,\cx)^*$,
determined by the $Ad^*$-invariant metric
$$
<X,Y>=\frac 1{2\pi i} \oint_{S^1}
\frac{tr(X(\lbd) Y(\lbd))} \lbd d\lbd,\ X,Y \in \wt{\grs\grl}(r,\cx).
\tag1.3
$$
Putting appropriate invariant restrictions on the leading terms of the
Laurent expansion of $\CalL(\lbd)$, the equations \thetag{1.1} become the
compatibility
conditions for the $x$- and $t$-flows determined by \thetag{1.2a,b}.

For a subalgebra $\grg \subset \grs\grl(r,\cx)$ with loop algebra
$\tilde \grg$ we denote by $\tilde \grg^-_0 \subset \tilde\grg$
the subalgebra of
elements extending holomorphically outside $S^1$,
identified
via $<\ , \ >$ with $\tilde\grg^{+*}$. (No notational distinction will be made
between elements of $\tilde\grg$ (resp. $\tilde\grg^-_0$) and the
corresponding element of $\tilde\grg^*$ (resp. $\tilde\grg^{+*}$).)
The specific form for $\Cal L$ giving rise to
equation \thetag{1.1} as compatibility conditions of (1.2a,b) is:
$$
\CalL(\lbd)=\lo+\lbd^{-1}\li+\lbd^{-2}\lii+\dots+\lbd^{-n+1}
\CalL_{n-1}\tag1.4
$$
with
$$
\align
\lo&=\frac{\r1}{a+b}\mx bI_a & 0 \\ 0 & -aI_b \emx \tag1.5a\\
\li &= \mx 0 & q \\ p & 0 \emx  \tag1.5b\\
\lii&=\r1\mx qp & -q_x \\ p_x & -pq \emx. \tag1.5c
\endalign
$$

A general framework for studying the ``finite-gap'' quasi-periodic solutions of
such Lax pair AKS flows was developed in \cite{AHP, AHH1, AHH2}, using
moment map embeddings of finite dimensional symplectic
vector spaces, reduced by certain Hamiltonian group
actions, into the duals of loop algebras. The image
space is a finite dimensional Poisson submanifold of $\wt{\grs\grl}(r,\cx)^*$
consisting of a union of  coadjoint orbits whose elements are rational
in the loop parameter. The original space, which we refer to
as the ``generalized Moser space'' (cf. \cite{AHP, M}), may be viewed as
consisting of pairs $(F,G)$ of maximal rank $(N\times r)$ matrices $(N>r)$
parametrizing rank-$r$ perturbations of
a fixed diagonal $N\times N$ matrix $A$, with eigenvalues
$\{\alpha_i\}_{i=1, \dots n}$ of multiplicities $\{k_i\}_{i=1,
\dots, n}$ and $k_i \le r,\,\sum_{i=1}^n k_i=N$. The general
form for such a moment
map is:
$$
\align
\Jtil:M^{N \times r} \times M^{N \times r}
&\longrightarrow  \wt{\grs \grl}(r,\cx)^{+*} \tag 1.6\\
(F, G) &\longmapsto \Cal N, \tag 1.6a
\endalign
$$
where
$$
\CalN(\lambda) = \lambda G\tp (A - \lambda I) F  ,\tag 1.6b
$$
$F$ and $G$ are $N \times r$ matrices chosen so that
$\tr(\CalN)=0$, and the space of pairs $(F,G)$ is given the
symplectic structure $$
\omega = \tr (dF\tp \wedge dG). \tag1.7
$$
We denote by ${\Cal M}^{\text{\bf k}}$, where $\text{\bf k}=(k_1,
k_2, \dots k_n)$, the space of pairs $(F,G)$ of such $N\times r$
matrices, subject to the generic requirement that the $k_i \times r$
dimensional blocks $F_i$ and $G_i$ corresponding to the eigenspaces of $A$ with
eigenvalue $\ai$
have maximal rank $k_i$, and endowed with the  symplectic structure
\thetag{1.7}.
Assuming $A$ to be diagonal, we obtain $\Cal N$ in the form:
$$
\Cal N(\lbd) = \sum_{i=1}^n \frac{\lbd {\Cal N}_i} {\alpha_i-\lbd}, \tag 1.8
$$
where
$$
{\Cal N}_i = G_i\tp F_i.
$$

Taking the AKS Hamiltonians
$\vPhi_x,\,\vPhi_t \in \CalF$
for the $x$ and $t$ flows as
$$
\align
\vPhi_x(X) &= \frac 12
\tr(\frac{a(\lbd)}{\lbd^{n-1}}X(\lbd)^2)_0 \tag{1.9a}\\
\vPhi_t(X) &= \frac 12
\tr(\frac{a(\lbd)}{\lbd^{n-2}}X(\lbd)^2)_0 \ ,  \tag{1.9b}
\endalign
$$
where the subscript $0$ means taking the $\lambda^0$ term in the Laurent
expansion centered at $\lbd=0$ and
$a(\lbd) = \prod_{i=1}^n (\lbd - \ai)$ is the minimal polynomial of
$A$, gives Hamilton's equations in the Lax pair form
$$
\align
\frac d{dx}\CalN(\lbd) &=
[(\frac{a(\lbd)}{\lbd^{n-1}}X(\lbd))_+,\CalN(\lbd)] \tag{1.10a}\\
\frac d{dt}\CalN(\lbd) &=
[(\frac{a(\lbd)}{\lbd^{n-2}}X(\lbd))_+,\CalN(\lbd)]. \tag{1.10b}
\endalign
$$
Defining
$$
\Cal L := -\frac{a(\lambda)}{\lambda^n} \Cal N(\lambda)\tag{1.11}
$$
gives the equations (1.2a,b), with $\CalL$ a polynomial in
$\lambda^{-1}$ of degree $n-1$. From equation \thetag{1.11}, we obtain
the leading terms  $\lo$, $\li$ and $\lii$ in terms of the matrices $(F_i,G_i)$
as
$$
\align
\lo&=\sn G_i\tp F_i \tag{1.12a}\\
\li&=\sn \ai (G_i\tp F_i - \lo) \tag{1.12b} \\
\lii&=-\sum_{j\geq k} \aj\alpha_k\lo + \sn \ai(\ai G_i\tp F_i - \li).
\tag{1.12c}
\endalign
$$

The form \thetag{1.5a} for the leading term $\lo$ determines an
invariant manifold under such AKS flows, since $\lo$ is just the
moment map generating
the conjugation action under $Sl(r,\cx)$, and hence is
conserved by the AKS flows. The other
two conditions \thetag{1.5b,c} determining the form of $\li$ and
$\lii$ are also invariant under the flows.

In the following
section we shall first briefly recall how to obtain the special
structure \thetag{1.5a-c} for $\CalL$ and
then show how the moment map \thetag{1.6} gives coordinates
parametrizing certain solutions of the nonlinear Schr\"odinger
equations associated to the Hermitian symmetric Lie algebra
$(\grs\grl(a+b,\cx),\grs\grl(a,\cx) \oplus \grs\grl(b,\cx) \oplus \cx^*)$
and its real
forms $(\grs\gru(a+r,s),\grs(\gru(a) \oplus \gru(r,s)))$. A Darboux coordinate
atlas is obtained for the reduced space $\Mk_{\grs\grl(a+b,\cx)}$, obtained
by requiring
$\CalN(\lbd)$ to be traceless, as well as for the real  forms obtained by
choosing fixed points under antilinear involutions.

Section 3 deals with hermitian symmetric pairs of the type
$( \grs\gro (2l),\gru(r) )$, $(\grs\gro (2l+2), \grs\gro(2l)
\oplus \grs\gro(2) )$, $( \grs\gro (2l+3) , \grs\gro(2l+1)
\oplus \grs\gro(2) )$ and $( \grs\grp(l), \gru(l) )$
as well as their corresponding noncompact and complex forms.
Global Darboux coordinates are
obtained from the
restriction that the moment map take its values in the total space of one of
the classical Lie algebras. The detailed
computations for these are omitted, the results being given in
tabular form.

We emphasize that the purpose of this work is to give an
intrinsic canonical parametrization of the invariant finite dimensional
sector of the phase space underlying certain solutions of the
matrix NLS equations;namely, those of ``finite gap'' type or more
generally, quasi-periodic flows lying on the stationary manifolds
of higher AKS invariants. The actual integration of
these finite dimensional flows in terms of theta functions, obtained from
linear flows of divisors on the
associated invariant spectral curves, will be the subject of a
subsequent work \cite{W}.

\bigskip
\noindent {\bf 2. Darboux Coordinates for the $(\grs\grl(a+b,\cx),
\grs\grl(a,\cx) \oplus \grs\grl(b,\cx) \oplus \cx)$
and \newline $(\grs\gru(a+l,s) , \grs(\gru(a) \oplus \gru(l,s)))$
Matrix NLS Equations}
\medskip
In this section, we compute Darboux coordinates for matrix nonlinear
Schr\"o\-din\-ger equations  of the type \thetag{1.1} and the real forms
$$
\r1 u_t = u_{xx} + 2(uD\bar{u}\tp)u , \tag{2.1}
$$
obtained by setting
$u=q=-D\bar q\tp$ where $u \in \cx^{a\times b}$ and $D$ is a Hermitian matrix
which, up to a base change can be taken
of the form $\smx I_l & 0 \\ 0 & -I_s \emx,\ l+s=b$.

First we derive the form for $\lo$, $\li$ and $\lii$ given in
\thetag{1.5a-c}. The co-adjoint action of the subgroup of
constant loops $Sl(r,\cx)$ on $\wt
{\grs\grl}(a+b,\cx)^{+*} \sim \wt{\grs\grl}(a+b,\cx)^-_0, \, a+b=r$, given by
$$
\Ad_g (X)(\lbd)=g(X(\lbd))g^{-1}, \tag2.2
$$
with $X(\lbd)=\sum_{i=-\infty}^0 X_i\lbd^i$,
 is generated by the moment map $J(X)=X_0$. Since our
Hamiltonians \thetag{1.9a,b} (and all other Hamiltonians in $\CalF$) are
invariant under this action,  $\lo$ is an invariant of the flows.

By a Hermitian symmetric Lie algebra (cf. \cite{KN}), we understand a Lie
algebra $\grg$ with subalgebra $\grk$ and an involutive automorphism
$\sig:\grg \rightarrow \grg$ such that $\grk$ is the $+1$ eigenspace and,
denoting by $\grm$ the $-1$ eigenspace, an element $B\in\grk$ exists
satisfying
$$
\align
i)&\ \grk=\text{ker}(ad\,B) \tag2.3a\\
ii)&\ (ad\,B|_\grm)^2=-\text{id}_\grm. \tag2.3b
\endalign
$$
The element $B$
has the the same properties \thetag{2.3a,b} in the complexification of $\grg$,
$\grk$ and $\grm$.
The $\grk$ and $\grm$ components of any element $X \in \grg$ will henceforth
be denoted $X_\grk$ and $X_\grm$ respectively.

In particular, for $\grg=\grs\grl(r,\cx)$ define the involutive automorphism
$$
\aligned
\s_0: \grs\grl(a+b,\cx) &\longmapsto \grs\grl(a+b,\cx)\\
X &\lra I_{a,b} X I_{a,b}
\endaligned\tag2.4
$$
with $I_{a,b}=\smx I_a & 0 \\ 0 & -I_b \esmx$. This also determines an
involutive automorphism $\tilde\s_0$
on the associated loop algebra $\tilde\grg$ through
$$
(\tilde\s_0(X))(\lbd)=\s_0(X(\lbd)),\ X\in\tilde\grg. \tag2.5
$$
The decomposition of $\grs\grl (a+b,\cx)$ into the $+1$ eigenspace $\grk=
\grs\grl(a,\cx) \oplus \grs\grl(b,\cx) \oplus \cx$ consisting of the diagonal
$a \times a$ and $b \times b$ blocks and the complementary
$-1$ eigenspace $\grm$ can be interpreted as the complexification of the
Hermitian symmetric Lie algebra $ (\grs\gru(a+b), \grs( \gru(a) \oplus
\gru(b) )$, with  properties \thetag{2.3a,b} in both cases satisfied by the
same element $B$.
Using the induced involution \thetag{2.5} we also have a
splitting of the loop algebra $\wt {\grs\grl} (a+b,\cx)$, with $+1$ eigenspace
$\tilde\grk$ and $-1$ eigenspace $\tilde\grm$ and
relations \thetag{2.3a,b} also hold when $\grk$, $\grm$ are replaced by
$\tilde\grk$, $\tilde\grm$.

Returning to the general case, since $\lo$ is an invariant of the
flows, we may choose it equal to $B\in\grk$, as mentioned above ($\grk$ is
considered here as representing the constant loops in $\tilde\grk$).
For any Hamiltonian $\phi \in \CalF$ with flow parameter $\tau$
let $d\phi(\CalL)_-=A\lbd^{-1}+\text{\bf O}(\lbd^{-2})$,
with $\CalL$ given by \thetag{1.4}. The AKS theorem implies
$$
\frac d{d\tau}\li=-[A,\lo].\tag{2.6a}
$$
It follows from \thetag{2.3a} that $\li{}_\grk$ is an invariant of the flows,
which we set equal to
zero as an initial condition. In particular, from Hamilton's equation
\thetag{1.2a}  we  deduce
$$
\frac d{dx}\li=[\lo,\lii] \in \grm \tag{2.6b}
$$
The equations of motion then also imply
$$
\lii{}_\grm = \frac d{dx}\li \tag2.6c
$$
due to \thetag{2.3b} and \thetag{2.6b}. Finally, an easy computation shows that
the condition
$$
\lii{}_\grk=\frac 12[[\lo,\li],\li] \tag2.6d
$$
is invariant under the flows of all the Hamiltonians in the
ring of functions $\CalF$, and hence may be  consistently imposed. For the
particular case $\grg=\grs\grl(r,\cx)$, we have $\lo=B$ given by equation
\thetag{1.5a} and equations \thetag{2.6a-d} are equivalent to the special form
\thetag{1.5b,c} for $\li,\,\lii$.

The moment map $\Jtil$ in \thetag{1.6} generally takes its values in
$\wt{\grg\grl}(r,\cx)^{+*}$. In order that it be defined so as to take
values in $\wt{\grs\grl}(a+b,\cx)^{+*}$, we have to impose restrictions
on its domain in $\Mk$.
The corresponding submanifold of $\Mk_{{\grs\grl}(r,\cx)} \subset \Mk$
is defined by
$$
\Mk_{{\grs\grl}(r,\cx)}=\{(F,G)\in\Mk\,|\,\tr(G_i\tp
F_i)=0,\,i=1,\dots,n\}.
$$
This is not a symplectic manifold; to obtain one, we have to quotient
by the $(\cx^*)^n$ group-action on $\Mk_{{\grs\grl}(r,\cx)}$
defined by $(z(F,G))_i=(z_i F_i,z_i^{-1} G_i),\,i=1,\dots,n$, where
$z=(z_1,\dots,z_n)\in(\cx^*)^n$, whose orbits are the null foliation of the
restriction of the symplectic form $\omega$ to ${\Mk_{{\grs\grl}(r,\cx)}}$.
This
amounts to a zero moment map reduction with respect to the action of the
central subgroup $\CalD$ of the loop group $\wt{Gl}(r,\cx)$ corresponding to
multiples of the identity matrix. Let $\pi:\Mk_{ {\grs\grl} (r,\cx) }
\rightarrow \Mk_{{\grs\grl}(r,\cx)}/(\cx^*)^n$ be the associated projection.
On $\Mk_{{\grs\grl}(r,\cx)}/(\cx^*)^n$ we have symplectic charts
$(\varphi^{\ltil\ctil},\pi(U^{\ltil\ctil}))$ defined as follows:
$$
\aligned
\varphi^{\ltil\ctil}\,:\,\pi(U^{\ltil\ctil}) &\longrightarrow
\varphi^{\ltil\ctil} ( \pi (U^{\ltil\ctil}) )\subset \cx^{Nr-n}
\times \cx^{Nr-n} \\
\pi (F,G) &\longmapsto (G\ce_i \tilde F_i,(G\ce_i)^{-1} \tilde
G_i)_{i=1, \dots ,n}
\endaligned\tag2.7
$$
where $\tilde F_i$ ($\tilde G_i$) means just $F_i$ ($G_i$) with the $F\ce$
($G\ce$) component suppressed, considered as an element of $\cx^{rk_i-1}$ and
$$
\gathered
U^{\ltil\ctil}=\{\pi (F,G)\in\Mk_{\wt{\grs\grl}(r,\cx)}/(\cx^*)^n \mid
G_i\ce\neq 0,\,F_i\ce\neq 0\} \\
\ltil=(l_1,\dots,l_n)\quad\ctil=(c_1,\dots,c_n)\\
l_i\in\{K_i+1, \dots, K_i+ k_i\} \quad
K_i:= \sum_{j=1}^{i-1} k_j \quad  c_i\in\{1,\dots,r\}
\endgathered\tag2.8
$$
But $\pi (U^{\ltil\ctil})$ is diffeomorphic to the following  symplectic
submanifold of $\Mk$:
$$
\pi (U^{\ltil\ctil}) \cong \{(F,G)\in\Mk \mid G_i\ce=1 , \
F_i\ce =-\sum_{(\da,\be) \ne (l_i,c_i)} G\dbi F\dbi \neq 0 ,\,i=1,\dots n\}
\tag2.9
$$
which defines a section of the quotient projection $\pi:\Mk_{ {\grs\grl}
(r,\cx) }
\rightarrow \Mk_{{\grs\grl}(r,\cx)}/(\cx^*)^n$.
The symplectic form \thetag{1.7} restricted to this manifold becomes
$$
\omega_{\pi(U^{\ltil\ctil})}=\sn
\undersetbrace{(\da ,\be ) \ne (l_i,c_i)} \to
{\sum_{\da=K_i+1}^{K_i+ k_i} \sum_{\be
=1}^r} dG\dbi\wedge dF\dbi.\tag2.10
$$
The moment map $\Jtil$ then is defined on the submanifolds \thetag{2.9} as
follows:
$$
\tilde J_{\grs\grl(a+b,\cx)} (\pi (F,G)) =\lbd \sn \frac
{G_i\tp F_i}{\al- \lbd} =-\frac {\lbd^n}{a(\lbd)}\sum _{i=0}^{n-1} \lbd^{-i}
\Ln_i \tag2.11
$$
where the map may be viewed as defined either on the quotient space
$\Mk_{{\grs\grl}(r,\cx)}/(\cx^*)^n$ or, equivalently, on the sections whose
images are defined by the diffeomorphism \thetag {2.9}.

Now, the condition \thetag{1.5a} imposed on $\lo$ pulls back
through the moment map to the space $\Mk$ and gives us the following level sets
on our charts $\pi (U^{\ltil\ctil})$ \thetag{2.9}.
Setting $F_i =(\xc,\xv)$ and  $G_i =(\yc,\yv)$ with $\xc, \yc\in \cx ^{k_i
\times a}$ and  $\xv, \,\yv \in \cx ^{k_i \times b}$ we obtain from the
diagonal blocks of the condition \thetag{1.5a}
$$
\align
\sn\yc\tp\xc&=\frac {b\r1}{a+b} I_a \tag2.12a\\
\sn\yv\tp\xv&=-\frac {a\r1}{a+b} I_b \ , \tag2.12b
\endalign
$$
and from its off-diagonal blocks
$$
\align
&\sn\yc\tp\xv=0 \tag2.13a\\
&\sn\yv\tp\xc=0. \tag 2.13b
\endalign
$$
The vanishing of the diagonal blocks in condition \thetag{1.5b} implies
$$
\align
\sn\al(\yc\tp\xc-\frac {b\r1}{a+b} I_a) &= 0\tag2.14a \\
\sn\al(\yv\tp\xv+\frac {a\r1}{a+b} I_b) &= 0,\tag2.14b
\endalign
$$
while the off-diagonal blocks give $q$ and $p$ in terms of the linear
coordinates \lb $(\xv,\xc,\yv,\yc)$ as
$$
q=\sn\al\yc\tp\xv, \qquad
p=\sn\al\yv\tp\xc.
\tag2.15
$$
Finally, the diagonal blocks of conditions \thetag{1.5c} imply the constraints
$$
\align
\sn\al^2\yc\tp\xc-\r1(qp+\sum_{j \geq k}\aj \ak \frac {b}{a+b} I_a)&=0
\tag2.16a\\
\sn\al^2\yv\tp\xv+\r1(pq+\sum_{j \geq k}\aj \ak \frac {a}{a+b} I_b)&=0.
\tag2.16b
\endalign
$$

Now let us turn to the general real form \thetag{2.1} of the nonlinear
Schr\"o\-din\-ger equation. The equations of motion and the choice of $\lo$,
$\li$ and $\lii$ are obtained in the same way as for the complex case,
with the understanding that we are now working in a reduction to the
Hermitian symmetric Lie algebra $(\grg=\grs\gru(a+l,s),\grk=\grs(\gru(a) \oplus
\gru(l,s)))$. This space is obtained by requiring invariance of the image of
the moment map under the antilinear involutive automorphism
$$
\aligned
\tilde \rho:\,\wt{\grs\grl}(a+l+s,\cx) &\rarrow \wt{\grs\grl}(a+l+s,\cx)\\
\tilde \rho (X)&= -t \Xbar\tp(\bar\lbd) t^{-1},
\endaligned\tag2.17
$$
where $t=\smx I_{a+l} & 0 \\ 0 & -I_s \esmx$ , as well as under the involution
\thetag{2.5}. Requiring that  the moment map
$\Jtil_{\grs\grl(a+b,\cx)}$  take its values in the fixed point
set $\widetilde{\grs\gru} (a+l,s)^* = \{ X \in \wt{\grs\grl}(a+l+s,\cx)^*
\mid \tilde \rho (X) = X \}$  forces further restrictions on the space
$\Mk_{{\grs\grl}(r,\cx)}$.  The method used in computing these
restrictions will apply equally to the reductions considered in Section 3.
(See also \cite{AHP} for details on the general approach to reductions under
involutive automorphisms). To implement such a discrete reduction we must
impose the following condition, reducing the domain of $\Jtil$ within the
space $\Mk_{{\grs\grl}(r,\cx)}$:
$$
\Wtil_{{\grs\gru}(a+l,s)}=\{\pi(F,G)
\in \Mk_{{\grs\grl}(r,\cx)}/(\cx^*)^n \mid \tilde\rho (\lbd \sn {G_i\tp
F_i\over \lbd-\al}) = \lbd \sn {G_i\tp F_i \over \lbd-\al}\, \quad \forall
 \lbd\}.
\tag2.18
$$
This implies that the $\ai$ are either real or come in complex conjugate
pairs. We may reorder the $\ai$ such that $\adi=\albar_{2i-1},
\ i=1, \dots, m$ and
$\aj=\ajb,\,j= 2m+1,\dots,n$. Thus \thetag{2.18} is equivalent to
 the constraints
$$
\align
G\tp_{2i} F_{2i} &= - t \Fbar\tp_{2i-1} \Gbar_{2i-1} t^{-1},\quad i=1, \dots,
m \tag2.19a\\
G\tp_j F_j &= -t \Fbar\tp_j \Gbar_j t^{-1} ,\quad j=2m+1, \dots, n\tag2.19b
\endalign
$$
 on
the pairs $(F_i,G_i)$.  The reduced space $\Mk_{{\grs\grl}(r,\cx)}/(\cx^*)^n$
splits
into several connected components satisfying these constraints, one of which is
$$
\aligned
W_{{\grs\gru}(a+l,s)} = \{\pi(F,G) \in \Mk_{{\grs\grl}(r,\cx)}/(\cx^*)^n
\mid
&G_{2i-1}=-\Fbar_{2i}t,\,
 G_{2i}=\Fbar_{2i-1}t,\ i=1,\dots,m,\\
&G_j=\r1\,\Fbar_j t,j=2m+1,\dots,n\ \}.
\endaligned\tag2.20
$$

Let $\pr_F:\,\cx^{Nr-n} \times \cx^{Nr-n} \rightarrow \cx^{Nr-n} $ be the
projection to the first factor. The charts
$(\varphi^{\ltil\ctil},\pi(U^{\ltil\ctil}))$, defined on
$\Mk_{{\grs\grl}(r,\cx)}/(\cx^*)^n$ give rise to charts
$(\varphi^{\ltil\ctil}_{  {\grs\gru}  },  U^{\ltil\ctil}_{
{\grs\gru} })$ on $W_{{\grs\gru}(a+l,s)}$ defined by
$$
\aligned
U^{\ltil\ctil}_{  {\grs\gru} } &= \pi (U^{\ltil\ctil}) \cap
W_{{\grs\gru}(a+l,s)} \\
\varphi^{\ltil\ctil}_{  {\grs\gru}  }&=
\pr_F \circ \varphi^{\ltil\ctil}|_{U^{\ltil\ctil}_{  {\grs\gru} }}
\endaligned\tag2.21
$$
which, in coordinates, gives:
$$
\varphi^{\ltil\ctil}_{  {\grs\gru}  } ( \pi(F,G) ) =
(G^{l_1,c_1}_1 \Ftil_1,\dots,G^{l_n,c_n}_n \Ftil_n) \in \cx^{Nr-n}.
\tag 2.22
$$
Restricting the symplectic form \thetag{2.10} to these charts we get the
following symplectic form on our new charts
$$
\aligned
\omega_{U^{\ltil\ctil}_{  {\grs\gru} }}=  \sm
&\left(
\undersetbrace {(\be,\da) \ne (l_i , c_i)} \to
{\sum_{\be=K_i + 1}^{K_i+k_i}
\sum_{\da=1}^{a+l}}(d\Fbar^{\be\da}_{2i} \wedge
dF^{\be\da}_{2i-1} + dF^{\be\da}_{2i}\wedge
d\Fbar^{\be\da}_{2i-1})\right. \\
&\left. -\undersetbrace {(\be,\da) \ne (l_i , c_i)} \to
{\sum_{\be=K_i + 1}^{K_i+k_i}\sum_{\da=a+l+1}^{a+l+s}}
(d\Fbar^{\be\da}_{2i}\wedge
dF^{\be\da}_{2i-1} + dF^{\be\da}_{2i}\wedge
d\Fbar^{\be\da}_{2i-1})\right) \\
 +&\r1 \sum_{j=2m+1}^n
\undersetbrace {(\bj,\dj) \ne (l_j , c_j)} \to
{\sum_{\be=K_i + 1}^{K_i+ k_i}
\sum_{\dj=1}^{a+l}} dF^{\bj\dj}_j\wedge d\Fbar^{\bj\dj}_j \\
-&\r1 \sum_{j=2m+1}^n
\undersetbrace {(\bj,\dj) \ne (l_j , c_j)} \to
{\sum_{\be=K_i + 1}^{K_i+ k_i} \sum_{\dj =a+l+1}^{a+l+s}}
dF^{\bj\dj}_j \wedge d\Fbar^{\bj\dj}_j) .
\endaligned\tag2.23
$$
Note that the terms with the minus sign are a consequence of the choice of the
matrix $t$ in \thetag{2.17}, \thetag{2.19a,b} and \thetag{2.20}.

Expressing the restriction of the moment map
$\Jtil_{{\grs\grl} (a+b,\cx)}$ to  $W_{{\grs\gru}(a+l,s)}$ as
$\Jtil_{{\grs\gru} (a+l,s)}$ we have:
$$
\spreadlines{3pt}
\aligned
\tilde J_{ {\grs\gru} (a+l,s)}
(\pi (F,G)) =
\lbd &\sm  \frac {\mx -\xdcb\tp\xuc & -\xdcb\tp\xuv \\
-D\,\xdvb\tp\xuc & -D\,\xdvb\tp\xuv \emx}{\adiu-\lbd} \\
&+ \lbd \sm\frac  {\mx
\xucb\tp\xdc & \xucb\tp\xdv \\
     D\,\xuvb\tp\xdc &
D\,\xuvb\tp\xdv \emx} {\adiub-\lbd}  \\
&+\r1\,\lbd  \sum_{j=2m+1}^n
\frac {\mx \xjcb\tp\xjc & \xjcb\tp\xjv \\
D\,\xjvb\tp\xjc &
D\,\xjvb\tp\xjv \emx}{\aj-\lbd}.
\endaligned\tag2.24
$$

The restrictions \thetag{1.5a-c} on $\lo,\li$ and $\lii$ required to obtain
the CNLS equation may be expressed in terms of the reduced coordinates over
the  space $W_{\grs\gru(a+l,s)}$. The constraints implied
by the choice \thetag{1.5a} of $\lo$ or equivalently \thetag{2.12a,b},
\thetag{2.13a,b} are
$$
\spreadlines{2pt}
\align
\sm (-\xdcb\tp\xuc + \xucb\tp\xdc) + \r1\sum_{j=2m+1}^n \xjcb\tp\xjc
&= \frac {b\,\r1}{a+b} I_a \tag2.25a\\
\sm (-\xdvb\tp\xuv + \xuvb \tp\xdv) + \r1\sum_{j=2m+1}^n \xjvb \tp\xjv
&= \frac {-a\,\r1}{a+b} D \tag2.25b \\
\sm (-\xdcb\tp\xuv + \xucb\tp\xdv) +
\r1\sum_{j=2m+1}^n \xjcb\tp\xjv &= 0.\tag2.25c
\endalign
$$
{}From the equations \thetag{2.15}, together with the reality conditions
\thetag{2.19a,b}, we find $u$ in terms of the reduced coordinates
$(\xc, \xv)$:
$$
u=\sm (\overline {\alpha}_{2i-1} \xucb\tp\xdv
                -\alpha_{2i-1}\xdcb\tp\xuv)
            +\r1 \sum_{j=2m+1}^n \aj \xjcb\tp\xjv. \tag2.26
$$
{}From the constraints \thetag{2.14a,b}, together with the reality conditions
\thetag{2.19a,b} we have
$$
\gather
\sm (-\alpha_{2i-1} \xdcb\tp\xuc
      + \albar_{2i-1} \xucb\tp\xdc)
+\r1(\sum_{j=2m+1}^n  \xjcb\tp\xjc - \sn\al\frac{b}{a+b}I_a)=0
\tag2.27a\\
\sm (-\alpha_{2i-1} \xdvb\tp\xuv
      + \albar_{2i-1} \xuvb\tp\xdv)
+ \r1(\sum_{j=2m+1}^n  \xjvb \tp\xjv + \sn\al\frac{a}{a+b}D)=0,\tag2.27b
\endgather
$$
while from the constraints \thetag{2.16a,b}, we have
$$
\align
\r1 \left( \sum_{j=2m+1}^n \alpha_j^2 \xjcb\tp \xjc
- \sum_{j \geq k} \aj \alpha_k {b \over {a+b}} I_a + uD\ubar\tp \right)
+ \sm (&-\alpha_{2i-1}^2 \xdcb\tp \xuc\\ &+ \albar_{2i-1}^2
\xucb\tp \xdc) = 0 \tag2.28a\\
\r1 \left( \sum_{j=2m+1}^n \alpha_j^2 \xjvb\tp \xjv
+ \sum_{j \geq k} \aj \alpha_k {a \over {a+b}} D - \ubar\tp u \right)
+ \sm (&-\alpha_{2i-1}^2 \xdvb\tp \xuv\\ &+ \albar_{2i-1}^2
\xuvb\tp \xdv) = 0. \tag2.28b
\endalign
$$

Thus, under the Hamiltonian flow induced by the collective Hamiltonians \lb
$\tilde J^*_{ {\grs\gru} (a+l,s)}\Phi_x$ and
$\tilde J^*_{ {\grs\gru} (a+l,s)}\Phi_t$ on $W_{\grs\gru(a+l,s)}$,
with $\Phi_x,\Phi_t$ given in equations \thetag{1.9a,b}, subject to the
invariant constraints \thetag{2.25a-c}, \thetag{2.28a,b}, the resulting
function
$u(x,t)$ given in equation \thetag{2.26} satisfies the CNLS equation
\thetag{2.1}.

\bigskip
\noindent {\bf 3. Symmetric Space Nonlinear Schr\"odinger Equations}
\medskip
The preceding section explained in detail how the
Lax pair flows determined by \thetag{1.10a,b} on $\wt{\grs\grl}(r,\cx)^{+*}$
give the reduced
Darboux coordinates and the invariant level sets on the reduced space
give rise to solutions of the coupled nonlinear Schr\"odinger (CNLS) equation
corresponding to the Hermitian symmetric Lie algebra
$(\grs\gru(a+l,s),\grs(\gru(a)\oplus \gru(l,s)))$ and its
complexification $(\grs\grl(a+b,\cx) , \grs\grl(a,\cx) \oplus \grs\grl(b,\cx)
\oplus \cx)$. In this section we construct reduced coordinates
for CNLS equations corresponding to the other Hermitian symmetric Lie algebras
considered by Fordy and Kulish \cite{FK}.
Referring to the classification of symmetric spaces in
\cite{H}, the Hermitian symmetric Lie algebras
$(\grs\gro(2l),\gru(l))\ (DIII)$ and $(\grs\grp(l),\gru(l))\ (CI)$
as well as their
corresponding noncompact and complex forms can be obtained as fixed points
of involutive automorphisms on $(\grs\gru(2l),\grs(\gru(l)\oplus\gru(l)))
\ (AIII)$ and on its corresponding noncompact and complex forms. On the other
hand, the complex structure for the $AIII$-cases is contained in the
$\gru(1)$-subalgebra, which is not invariant under the involution leading
to the Hermitian symmetric Lie algebras associated with
$(\grs\gro(2+k) , \grs\gro(2) \oplus \grs\gro(k))\ (BDI)$.
Therefore the $BDI$-Hermitian symmetric Lie algebras cannot be realized
as subcases of the $AIII$-cases. Nevertheless, we can obtain them by reduction
of the corresponding generic systems in $\grg\grl(r,\cx)$.

For all these cases the resulting Darboux coordinates are global and directly
obtained by a
reduction of $\grg\grl(r,\cx)$ to the classical complex Lie algebras $\grb_n$,
$\grc_n$, $\grd_n$ and to their corresponding real forms.
We could have obtained the coordinates for the $CI$- and $DIII$-cases by
imposing further constraints on the
coordinates associated with the $AIII$-cases, but the reduction from
$\grg\grl(r,\cx)$ avoids the need for local charts.
We emphasize that only the constraints on the Moser space induced by
reductions to the classical Lie algebras are solved, and not the additional
invariant constraints arising from the splitting $\grg=\grk \oplus \grm$ of the
Hermitian symmetric algebras underlying equations \thetag{1.5a-c}. The latter
are simply imposed, as in the previous section, upon the initial data, and
continue to hold valid under the flows.

The following are the involutions to be applied to the
$AIII$-Hermitian symmetric Lie algebras in order to get the $CI$- and
$DIII$-Hermitian
symmetric Lie algebras.
$$
\align
\sig_{CI}(X) &=  \smx 0 & I \\ -I & 0 \esmx
 X\tp \smx 0 & I \\ -I & 0 \esmx \tag3.1a \\
\sig_{DIII}(X) &= - \smx 0 & I \\ I & 0 \esmx X\tp
\smx 0 & I \\ I & 0 \esmx \tag 3.1b
\endalign
$$

The corresponding additional restrictions
on the functions $q,p$ and $u$ defined by equations \thetag{2.15} and
\thetag{2.26} and the equations \thetag{1.1} and
\thetag{2.1} are given in Table $I$. Note that in all these equations, the
matrices $p$, $q$ and $u$ are of dimension $l \times l$. The restrictions
given in rows 2 and 3 should be interpreted as invariant quadratic constraints
satisfied by the matrices $(F_i,G_i)$ on a suitably defined reduced manifold
$W$.
The general form of the matrices $\lo$, $\li$ and $\lii$ entering in the Lax
equations \thetag{1.2a,b} is still given by equations \thetag{1.5a-c}.

\medskip
%
%
\centerline{Table $I$}
\nopagebreak
\smallskip
\boxedalign
height3pt&\omit&&\multispan3&&\multispan3&&\multispan3&\cr
&
&& \multispan3\hfil\mdbox{(\grs\grl(2l,\cx),\grg\grl(l,\cx))}\hfil
&& \multispan3\hfil\mdbox{(\grs\gru(2l),\grs(\gru(l) \oplus \gru(l)))}\hfil
&& \multispan3\hfil\mdbox{(\grs\gru(l,l),\grs(\gru(l) \oplus \gru(l)))}\hfil
&\cr
height 3pt&\omit&&\multispan3&&\multispan3&&\multispan3&\cr
\noalign{\hrule height 0.8pt}
height 3pt&\omit&&\multispan3&&\multispan3&&\multispan3&\cr
& \vbox{\hbox{Correspon-} \vglue 2pt \hbox{ding CNLS} \vglue 2pt
\hbox{equation}}
&& \multispan 3\hfil
\mdbox{\aligned
\r1 q_t &= q_{xx}-2qpq \\
-\r1 p_t &= p_{xx}-2pqp
\endaligned}\hfil
&& \multispan 3\hfil
\mdbox{\r1 u_t=u_{xx} + 2u\ubar\tp u}\hfil
&& \multispan 3\hfil
\mdbox{\r1 u_t=u_{xx} - 2u\ubar\tp u}\hfil
&\cr
height 3pt&\omit&&\multispan3&&\multispan3&&\multispan3&\cr
\noalign{\hrule height 0.8pt}
\espho
& \vbox{\hbox{Involution:} \vglue 2pt \mdbox{\ \ \ \sig_{CI}}}
&& \mdbox{\frac{\grs\grp(2l,\cx)}{\grg\grl(l,\cx)}}
&& \mdbox{\aligned q&=q\tp \\ p&=p\tp \endaligned}
&& \mdbox{\frac{\grs\grp(l)}{\gru(l)}}
&& \mdbox{u=u\tp}
&& \mdbox{\frac{\grs\grp(l,\rx)}{\gru(l)}}
&& \mdbox{u=u\tp}
&\cr
\alargeline
& \vbox{\hbox{Involution:} \vglue 2pt \mdbox{\ \ \ \sig_{DIII}}}
&& \mdbox{\frac{\grs\gro(2l,\cx)}{\grg\grl(l,\cx)}}
&& \mdbox{\aligned q&=-q\tp \\ p&=-p\tp \endaligned}
&& \mdbox{\frac{\grs\gro(2l)}{\gru(l)}}
&& \mdbox{u=-u\tp}
&& \mdbox{\frac{\grs\gro^*(2l)}{\gru(l)}}
&& \mdbox{u=-u\tp}
&
\endboxedalign
%
%

\smallskip
For the Hermitian symmetric Lie algebras corresponding to $BDI$,
the Lie-algebras $\grs\gro(2+2l,\cx)$
and $\grs\gro(2+2l+1,\cx)$ are obtained as fixed point sets of the involutions
$$
\align
\sig_{DI}(X)&=-\smx 0 & I_{l+1} \\ I_{l+1} & 0 \esmx
X\tp \smx 0 & I_{l+1} \\ I_{l+1} & 0 \esmx,\, X \in
\grs\grl(2+2l,\cx)\tag 3.2a \\
\sig_{BI}(X)&=-\smx 1 & 0 & 0 \\ 0 & 0 & I_{l+1} \\ 0 & I_{l+1} & 0 \esmx
X\tp \smx 1 & 0 & 0 \\ 0 & 0 & I_{l+1} \\ 0 & I_{l+1} & 0 \esmx,\,
X \in \grs\grl(2+2l+1,\cx) \tag3.2b
\endalign
$$
(The separation into ``$BI$'' and ``$DI$'' is necessary due to our choice of a
basis that makes the relation between the Lax matrices in \thetag{1.5a-c}
and the corresponding nonlinear Schr\"odinger equations more transparent.)
The involutive automorphisms determining the Hermitian symmetric Lie algebra
decomposition are
$$
\align
\bar\sig_{DI}(X)&=\smx I_{1,l} & 0 \\ 0 & I_{1,l} \esmx
X\smx I_{1,l} & 0 \\ 0 & I_{1,l} \esmx,\quad X\in \grs\gro(2+2l,\cx) \tag 3.3a
\\
\bar\sig_{BI}(X)&=\smx -1 & 0 & 0 \\ 0 & I_{1.l} & 0 \\ 0 & 0 & I_{1,l} \esmx
X \smx -1 & 0 & 0 \\ 0 & I_{1.l} & 0 \\ 0 & 0 & I_{1,l} \esmx,\quad
X \in \grs\gro(2+2l+1,\cx). \tag3.3b
\endalign
$$
This gives the splitting $\grs\gro(2+2l,\cx)=\grk^1
 \oplus \grm^1$ and $\grs\gro(2+2l+1,\cx)=
\grk^2 \oplus \grm^2$ respectively, where $\grk^1:=
\grs\gro(2,\cx) \oplus \grs\gro(2l,\cx)$ and $\grk^2:=
\grs\gro(2,\cx) \oplus \grs\gro(2l+1,\cx)$ are the $+1$ eigenspaces. The
corresponding elements $\lo$, $\li$ and $\lii$ given by equations
\thetag{1.12a-c} will
be denoted $\lo^s$, $\li^s$ and $\lii^s,\ s=1,2$ respectively, for the two
cases. In the two
cases the complex structures $I^1:=ad{\lo^1}|_{\grm^1}$,
(resp. $I^2:ad{\lo^2}|_{\grm^2}$) underlying the Hermitian symmetric
Lie algebra is given by an element $\lo^1$ (resp. $\lo^2$)
of the $\grs\gro(2)$-subalgebra of the
$+1$ eigenspace of the involutions \thetag{3.3a,b}. Denoting by $E_{i,j}$
the matrix with $1$ in the $ij^{\underline{\hbox{\sevenrm th}}}$ position
and zero elsewhere, we set
$$
\align
\lo^1 &=\r1 E_{1,1} - \r1 E_{l+1,l+1} \in \grs\gro(2+2l,\cx) \tag 3.4a \\
\lo^2 &=\r1 E_{2,2} - \r1 E_{l+2,l+2} \in \grs\gro(2+2l+1,\cx). \tag 3.4b
\endalign
$$

Using the same techniques as in the generic case we see that, with
$\lo^1,\lo^2$
chosen as above, the $\grk^1$ (resp. $\grk^2$) components of $\li^1,\li^2$
are invarants
of the flows which we choose equal to zero. This gives the general form:
$$
\align
\li^1&=
\mx 0 & \ktmi q\tp & 0 &  \ktmi \qt\tp \\ p & 0 &
\kmmi -\qt & 0 \\ 0 & -\pt\tp & 0 & -p\tp \\
\pt & 0 & \kmmi -q & 0 \emx \in \grs\gro(2+2l,\cx) \tag 3.5a\\ \vspace{5pt}
\li^2&=
\mx 0 & -p^1 & 0 & \kenmi q^1 & 0 \\ \vs
         -q^1 & 0 & \ktmi q\tp & 0 & \ktmi\qt\tp \\ \vs
         0 & p & 0 & \kmmi -\qt & 0 \\ \vs
         \kenmi p^1 & 0 & -\pt\tp &  0 & -p\tp \\ \vs
         0 & \pt & 0 & \kmmi -q & 0 \emx
\in \grs\gro(2+2l+1,\cx) \tag 3.5b
\endalign
$$
where $q,\qt,p,\pt\in\cx^l$ and $q^1,p^1\in\cx$.
Then $\lii^1,\lii^2$ are obtained by substituting \thetag{3.5a,b} into
equations
\thetag{2.6c,d}.

{}From the compatibility conditions
for equations \thetag{1.2a,b} with $\lo^s,\,\li^s$ and $\lii^s,\,s=1,2$ chosen
as
above, we get the coupled nonlinear Schr\"odinger equations
$$
\aligned
\r1 q_t &= q_{xx} - 2q(q\tp p + \qt\tp\pt) + 2
q\tp\qt\pt \\
\r1 \qt_t &= \qt_{xx} - 2\qt(q\tp p +
\qt\tp\pt) + 2\qt\tp qp \\
 -\r1 p_t &= p_{xx} - 2p(q\tp p
+ \qt\tp\pt) + 2p\tp\pt\qt \\
-\r1 \pt_t &= \pt_{xx} -
2\pt(q\tp p + \qt\tp\pt)  + 2\pt\tp pq
\endaligned\tag 3.6a
$$
for the Hermitian symmetric Lie algebra $({\grs\gro(2+2l,\cx)} ,
\grs\gro(2,\cx) \oplus \grs\gro(2l,\cx))$,  and
$$
\aligned
\r1 q^1_t &= q^1_{xx} - 2 q^1(q\tp p + \qt\tp\pt + q^1 p^1)
+ p^1( q^1 q^1 + 2q\tp\qt) \\
-\r1 p^1_t &= p^1_{xx} - 2p^1(q\tp p + \qt\tp\pt+q^1 p^1)
+ q^1(p^1 p^1 + 2 p\tp\pt) \\
\r1 q_t &= q_{xx} - 2q(q\tp p + \qt\tp\pt + q^1 p^1)
+ \pt(q^1 q^1 + 2q\tp\qt) \\
-\r1 p_t &= p_{xx} - 2p(q\tp p + \qt\tp\pt + q^1 p^1)
 + \qt(p^1 p^1 + 2p\tp\pt) \\
\r1\qt_t &= \qt_{xx} - 2\qt(q\tp p + \qt\tp\pt + q^1 p^1)
+ p(q^1 q^1 + 2\qt\tp q) \\
-\r1\pt_t &= \pt_{xx} - 2\pt(q\tp p + \qt\tp\pt + q^1 p^1)
+ q(p^1 p^1 + 2\pt\tp p)
\endaligned \tag3.6b
$$
for the Hermitian symmetric Lie algebra
$(\grs\gro(2+2l+1,\cx) , \grs\gro(2,\cx)
\oplus \grs\gro(2l+1,\cx))$.

The reduction to the compact real forms $(\grs\gro(2+2l) ,
\grs\gro(2) \oplus \grs\gro(2l))$ and $(\grs\gro(2+2l+1) ,
\grs\gro(2) \oplus \grs\gro(2l+1))$ is induced by the involutive
automorphism
$$
\sig(X)=-\Xbar\tp\tag 3.7
$$
for both cases. The resulting reality conditions for $\li^1$ and $\lii^1$
which reduce equations \thetag{3.6a} are
$$
q=-\pbar \qquad \qtil=-\contil p. \tag 3.8a
$$
The reality conditions for $\li^2$, $\lii^2$ reducing equations
\thetag{3.6b} are
$$
q=-\pbar \qquad \qtil=-\contil p \qquad q^1=-\pbar^1. \tag 3.8b
$$

The noncompact real forms $(\grs\gro(2,2l),
\grs\gro(2) \oplus \grs\gro(2l))$ and $(\grs\gro(2,2l+1) ,
\grs\gro(2) \oplus \grs\gro(2l+1))$ are obtained through the involutive
automorphisms
$$
\align
\rho_{DI}(X) &= - \smx I_{1,l} & 0 \\ 0 & I_{1,l} \esmx \Xbar\tp
\smx I_{1.l} & 0 \\ 0 & I_{1,l} \esmx,\,X \in \grs\gro(2+2l,\cx) \tag3.9a\\
\rho_{BI}(X) &= - \smx -1 & 0 & 0 \\ 0 & I_{1,l} & 0 \\ 0 & 0 & I_{1.l} \esmx
\Xbar\tp \smx -1 & 0 & 0 \\ 0 & I_{1,l} & 0 \\ 0 & 0 & I_{1.l} \esmx,\,
X \in \grs\gro(2+2l+1,\cx) \tag3.9b
\endalign
$$
The resulting reduced form of $\li^1$, $\lii^1$ and equations \thetag{3.6a}
are given by
$$
q=\pbar \qquad \qtil=\contil p. \tag 3.10a
$$
The reality conditions giving the reduced form of $\li^2$, $\lii^2$ and
equations \thetag{3.6b} are
$$
q=\pbar \qquad \qtil=\contil p \qquad q^1=\contil p^1. \tag 3.10b
$$

\bigskip
Tables $II(a-c)$ give a list of reduced
coordinates, corresponding symplectic forms and
Lax matrices $\CalN(\lbd)$ for all the Hermitian symmetric Lie algebras
listed above. The reductions are given relative
to the generic case by expressing the pairs
$(F,G)\in \Mk$ of $N\times r$-complex matrices in terms of reduced coordinates,
denoted $(X,Y)$ or $(w,X,Y)$, where $X$ and $Y$ are reduced rectangular blocks
and $w\in \cx^N$, subject to the remaining invariant  constraints.
The second column gives the restrictions and reality conditions on the
$k_i\times r$-blocks $(F_i,G_i)$ as well as constraints on the eigenvalues of
$A$ required in order that the moment map take its values in
the subalgebras  $\tilde\grg^{+*} \subset \wt{\grs\grl}(r,\cx)^{+*}$
corresponding to
the various Hermitian symmetric Lie algebras $(\grg,\grk)$. The independant
variables entering in the $k_i\times r$ blocks $(F_i,G_i)$ are denoted
$(X_i,Y_i)$ or $(w_i,X_i,Y_i)$ for the various cases  as defined in the table.
The invariant constraints to which these reduced coordinates are subject are
obtained by substituting the expressions for $(F,G)$ given
in the second column of Tables $II(a-c)$ into equations
\thetag{1.12a-c} and imposing
the conditions \thetag{1.5a-c} for the cases in Table $I$ and \thetag{3.4a,b},
together with the vanishing terms in \thetag{3.5a,b} and the conditons
\thetag{2.6d} imposed
on $\lii^s\lower 3pt\msbox{\sst\grk}$ for the remaining cases $BI$ and $DI$.
The third column gives the corresponding
symplectic forms in terms of the reduced coordinates.
The fourth column expresses the matrices $q,\qt,q^1,p,\pt,p^1$ or $u,u^1,v$
satisfying the
matrix CNLS equations in terms of these reduced coordinates. The last column
expresses, in terms of the reduced coordinates, the image
$\CalN(\lbd)$ of the moment map whose isospectral flows,
determined by equations \thetag{1.2a,b},
imply that $q,\qt,q^1,p,\pt,p^1,u,u^1,v$ satisfy the CNLS equations,
given either in the first row of Table $I$ or in equations \thetag{3.6a} or
\thetag{3.6b}.

\newpage

\centerline{\bf References}
\smallskip
\item{\bf [AHH1]} Adams, M.R., Harnad, J. and Hurtubise, J.,
{\sl Isospectral Hamiltonian Flows in Finite and Infinite Dimensions II.
Integration of Flows},
Commun. Math. Phys. {\bf 134} (1990), 555--585.
\item{\bf [AHH2]} Adams, M.R., Harnad, J. and Hurtubise, J.,
{\sl Dual Moment Maps to Loop Algebras},
Lett. Math. Phys. {\bf 20} (1990), 294--308.
\item{\bf [AHP]} Adams, M.R., Harnad, J. and Previato, E.,
{\sl  Isospectral Hamiltonian Flows in Finite and Infinite Dimensions I.
Generalized Moser Systems and Moment Maps into Loop Algebras},
Commun. Math. Phys. {\bf 117} (1988), 451--500.
\item{\bf [FK]} Fordy, A.P. and Kulish, P.P.,
{\sl Nonlinear Schr\"odinger Equations and Simple Lie Algebras},
Commun. Math. Phys. {\bf 89} (1983), 427--443.
\item{\bf [H]} Helgason, S.,
``Differential Geometry, Lie Groups and Symmetric Spaces'', Academic Press,
New York, 1978.
\item{\bf [KN]} Kobayashi, S. and Nomizu, K.,
``Foundations of Differential Geometry, Vol. II'', John Wiley, Interscience,
New York, 1969.
\item{\bf [M]} Moser, J., {\sl Geometry of Quadrics and Spectral Theory},
in ``The Chern Symposium, Berkeley, june 1979'',
Springer Verlag, Berlin Heidelberg New York, 1980, pp. 147--188.
\item{\bf [W]} Wisse, M.A.,
{\sl Periodic Solutions of Matrix Nonlinear Schr\"odinger Equations}
to appear: J. Math. Phys. {\bf 33}(12), (1992).

\newpage
\NoPageNumbers
\hoffset=-42pt
\hsize=10 true in
\vsize=7.5 true in


\def\contil#1{\kern2pt\bar{\kern-2pt{\tilde #1}}}
\def\espho{height2pt&\omit&&\omit&&\omit&&
          \omit&&\omit&\cr}
\def\ncaliu{\CalN^{(1)}_i}
\def\ncalid{\CalN^{(2)}_i}
\def\ncalj{\CalN_j}
\def\ncali{\CalN_i}
\def\sst{\scriptstyle}
\def\dst{\displaystyle}
\def\espace{\hskip 3pt\relax}
\def\aldi{\alpha_{2i}}

\def\mdbox#1{\hbox{$\dst #1$}}
\def\msbox#1{\hbox{$\sst #1$}}
\def\separer{\vglue 4pt}

\def\textfrac#1#2{\vbox{\hbox{\hglue 12pt #1}\vglue
2pt\hrule \vglue 2pt\hbox{#2}}}
\def\boxedalign{\bgroup\offinterlineskip\vbox\bgroup
\hrule height 0.8pt
\ialign \bgroup &\vrule width 0.8pt##&
\espace\hfil$\vcenter{##}$\hfil\espace&\vrule width 0.8pt##&
\espace\hfil$\vcenter{##}$\hfil\espace&\vrule##&
\espace\hfil$\vcenter{##}$\hfil\espace&\vrule##&
\espace\hfil$\vcenter{##}$\hfil\espace&\vrule##&
\espace\hfil$\vcenter{##}$\hfil\espace&\vrule
width 0.8pt##\cr\espho }
\def\endboxedalign{\cr\espho\egroup\hrule height
0.8pt\egroup\egroup }

\centerline{\bf Table $II$a}
\centerline{Darboux Coordinates for CNLS Equations ($CI$ and $DIII$)}
\vglue 2pt
\boxedalign
&
\textfrac{Algebra}{+1 eigenspace}
&&\hbox{Reduced coords.}\vglue 2pt \hbox{\& reality conds.}
&&\hbox{Symplectic form}
&&\hbox{$p,q;u(X,Y)$}
&&\hbox{Moment map $\CalN(\lbd)$ residues$^{1,2}$}
&\cr
\espho
\noalign{\hrule height 0.8pt}
\espho
& \mdbox{\frac{{\grs\grp}(r,\cx)}{{\grg\grl}(r,\cx)}}
&& \mdbox{
\aligned
F&=(X,Y) \\
G&=(-Y,X) \\
F_i&=(X_i,Y_i) \\
G_i&=(-Y_i,X_i) \endaligned}
\vglue 2pt
\mdbox{X,Y \in \cx^{N \times r}}
\mdbox{X_i,Y_i \in \cx^{k_i \times r}}
&& \mdbox{2\,tr(dY\wedge dX\tp)}
&& \mdbox{
\aligned
q&=-\sn\al Y\tp_i Y_i\\
p&=\sn\al X\tp_i X_i\endaligned}
&&
\mdbox{\ncali=\smx -Y\tp_i
X_i & -Y\tp_i Y_i \\ X\tp_i X_i & X\tp_i Y_i\esmx}
&\cr
\espho
\noalign{\hrule}
\espho
&\mdbox{\frac{{\grs\grp}(r)}{\gru(r)}}
&& \halign{\hfil$\dst#$&$\dst#$\hfil\cr
\adiu&=\adib \cr
X_{2i-1}&=-\Ybar_{2i}\cr
Y_{2i-1}&=\Xbar_{2i}\cr}\vglue 2pt
\mdbox{i=1,\dots,m= n/2}
&&
\mdbox{ 2\sm tr(dY_{2i}\wedge dX_{2i}\tp + d\Ybar_{2i}\wedge
d\Xbar\tp_{2i}) }
&&
\mdbox{u=-\sm (\adib\Xbar_{2i}\tp\Xbar_{2i} +\aldi Y_{2i}\tp Y_{2i}) }
&&
\mdbox{
\aligned
\ncaliu&=\smx -\yd\tp \xd & -\yd\tp \yd \\ \xd\tp \xd & \xd\tp \yd
\esmx  \\
\ncalid&=\smx \Xbar_{2i}\tp\Ybar_{2i} & -\Xbar_{2i}\tp\Xbar_{2i} \\
\Ybar_{2i}\tp\Ybar_{2i} & -\Ybar_{2i}\tp\Xbar_{2i} \esmx
\endaligned}
&\cr
\espho
\noalign{\hrule}
\espho
&\mdbox{\frac{{\grs\grp}(r,\rx)}{\gru(r)}}
&& \mdbox{
\aligned
\aj&=\ajb \\
X_j&=-\r1\,\mu_j\Ybar_j \\
\mu_j&=I_{p_j,q_j} \\
p_j&=k_j-q_j \endaligned}\vglue 3pt
\mdbox{j=1,\dots,n;\,m=0}
&&\mdbox{-2\,\r1\,\sj tr(dY_j\wedge
d\Ybar_j\tp\mu_j)}
&&\mdbox{u=-\sj\aj Y_j\tp Y_j}
&&\mdbox{\ncalj =
\smx \r1\,Y_j\tp\mu_j\Ybar_j & -Y_j\tp Y_j \\
-\Ybar_j\tp\Ybar_j & -\r1\,\Ybar_j\tp\mu_j Y_j \esmx}
&\cr
\espho
\noalign{\hrule height 0.8pt \vglue 5pt\hrule height 0.8pt}
\espho
&\mdbox{\frac{{\grs\gro}(2r,\cx)}{\grg\grl(r,\cx)}}
&&\mdbox{G_i=(X_i,Y_i)}\vglue 2pt
\mdbox{F_i=\gki(Y_i,X_i)}\vglue 2pt
\mdbox{X_i,Y_i\in\cx^{2\kappa_i\times r}}
\vglue 2pt
\mdbox{i=1,\dots ,n}
&&\mdbox{2\sn tr(\gki( dX_i\wedge dY_i\tp))}
&&\mdbox{
\aligned
q &= \sn\al X_i\tp\gki X_i\\
p &= \sn\al Y_i\tp\gki Y_i
\endaligned}
&&\mdbox{\ncali = \smx X_i\tp\gki Y_i & X_i\tp\gki X_i \\
Y_i\tp\gki Y_i &  Y_i\tp\gki X_i\esmx}
&\cr
\espho
\noalign{\hrule}
\espho
&\mdbox{\frac{{\grs\gro}(2r)}{\gru(r)}}
&&\mdbox{
\aligned
\adiu&=\adib\\
Y_{2i-1}&=-\gkti\Xbar_{2i}\\
X_{2i-1}&=-\gkti\Ybar_{2i}
\endaligned}\separer
\mdbox{i=1,\dots,m}\separer\separer
\mdbox{
\aligned
\aj&=\ajb\\
Y_j&=-\r1\gkj\rj\Xbar_j
\endaligned}\separer
\mdbox{ j=2m+1,\dots,n}
&&
\mdbox{ 2\sm tr(\gkti
(d\xd\wedge d\yd\tp+d\xdb\wedge d\ydb\tp))} \vglue 3pt
\mdbox{ +2\r1 \sjm tr(\rj d\xjb\wedge dX_j\tp)}
&&
\mdbox{ u=\vtop{
\mdbox{\sm (\adi \xd\tp \gkti \xd+\adib \ydb\tp \gkti \ydb)}
\mdbox{\ + \sjm \aj X_j\tp \gkj X_j}}}
&&
\mdbox{
\aligned
\ncaliu &= \smx \xd\tp\gkti\yd & \kenmi\xd\tp\gkti\xd \\
 \yd\tp\gkti\yd & \kenmi\yd\tp\gkti\xd \esmx \\
\ncalid &= \smx \ydb\tp\gkti\xdb &  \kenmi\ydb\tp\gkti\ydb\\
\xdb\tp\gkti\xdb & \kenmi\xdb\tp\gkti\ydb \esmx\\
\ncalj &= \smx \r1\, X_j\tp\rj\xjb & X_j\tp\gkj X_j \\
\xjb\tp\gkj\xjb & -\r1\,\xjb\tp\rj X_j \esmx
\endaligned }
&\cr
\espho
\noalign{\hrule}
\espho
&\mdbox{\frac{{\grs\gro}^*(2r)}{\gru(r)}}
&&\mdbox{
\aligned
\aj&=\ajb \\
Y_j&=-\r1\gkj\rj\xjb \endaligned}\separer
\mdbox{ j=1,\dots,n;\,m=0}
&&
\mdbox{ 2\,\r1\,\sj tr(d\xjb\wedge dX_j\tp) }
&&
\mdbox{ u=\sj\aj X_j\tp\gkj X_j }
&&
\mdbox{ \ncalj =
\smx \r1\,X_j\tp\xjb & X_j\tp\gkj X_j \\
-\xjb\tp\gkj\xjb & -\r1\,\xjb\tp X_j \esmx}
&
\endboxedalign

\newpage
\centerline{\bf Table $II$b}
\centerline{Darbux Coordinates for CNLS Equations ($DI$)}
\vglue 2pt
\boxedalign
&
\textfrac{Algebra} {+1 eigenspace}
&&\hbox{Reduced coords.}\vglue 2pt \hbox{\&
 reality conds.}
&&\hbox{Symplectic form}
&&\hbox{$q,\qt,p,\pt;u,v(X,Y)$}
&&\hbox{Moment map $\CalN(\lbd)$ residues$^{1,2}$}
&\cr
\espho
\noalign{\hrule height 0.8pt}
\espho
&\mdbox{\frac{\grs\gro(2+2l,\cx)}
{\grs\gro(2,\cx) \oplus \grs\gro(2l,\cx)}}
&&
\mdbox{ G_i=(X_i,Y_i)}\separer
\mdbox{ F_i=\gki(Y_i,X_i)}\separer
\mdbox{ X_i,Y_i\in\cx^{2\kappa_i\times r}}\separer
\mdbox{i=1,\dots,n}
&&
\mdbox{2\sn tr(\gki(dX_i \wedge dY_i\tp))}
&&
\mdbox{\aligned
q &= -\sn\al(\Yhat_i)\tp\gki X_i^1 \\
\qt &= -\sn\al(\Xhat_i)\tp\gki X_i^1 \\
p &= \sn\al(\Xhat_i)\tp\gki Y_i^1 \\
\pt &= \sn\al(\Yhat_i)\tp\gki Y_i^1
\endaligned}
&&
\mdbox{\ncali= \smx X_i\tp\gki Y_i &  X_i\tp\gki X_i \\
Y_i\tp\gki Y_i & Y_i\tp\gki X_i\esmx}
&\cr
\espho
\noalign{\hrule}
\espho
&\mdbox{\frac{{\grs\gro}(2+2l) }
{\grs\gro(2)\oplus \grs\gro(2l)} }
&&
\mdbox{
\aligned
\adiu&=\adib\\
Y_{2i-1}&=-\gkti\Xbar_{2i}\\
X_{2i-1}&=-\gkti\Ybar_{2i}\endaligned}\separer
\mdbox{i=1,\dots,m}\separer
\mdbox{
\aligned
\aj&=\ajb\\
Y_j&=-\r1\gkj \rj\Xbar_j\endaligned}\separer
\mdbox{j=2m+1,\dots,n}
&&
\mdbox{2\sm
\vtop{
\mdbox{ tr(\gkti\vtop{
\mdbox{(d\xd\wedge d\yd\tp}\vglue 2pt
\mdbox{\  + d\xdb\wedge d\ydb\tp))}}}
\mdbox{\ +2\,\r1 \sjm tr(\rj d\xjb\wedge dX_j\tp)}}}
&&
\mdbox{\aligned
u &= -\sm \vtop{\mdbox{ ( \adi\yth\tp\gkti\xd^1}
\mdbox{\quad + \adib\xthb\tp\gkti\ydb^1)}
\mdbox{+\r1\sjm\aj\xjhb\tp\rj X_j^1}}\\
v &=-\sm \vtop{
\mdbox{(\adi \xth\tp \gkti \xd^1}
\mdbox{\ \ +\adib \ythb\tp \gkti \ydb^1)}
\mdbox{ +\sjm \aj \Xhat_j\tp \gkj X_j^1}}
\endaligned}
&&
\mdbox{
\aligned
\ncaliu &= \smx \xd\tp\gkti\yd & \kenmi\xd\tp\gkti\xd \\
 \yd\tp\gkti\yd &  \kenmi\yd\tp\gkti\xd \esmx \\
\ncalid &= \smx\ydb\tp\gkti\xdb & \kenmi\ydb\tp\gkti\ydb \\
\xdb\tp\gkti\xdb & \kenmi\xdb\tp\gkti\ydb \esmx  \\
\ncalj &= \smx \r1 X_j\tp \rj\xjb &  X_j\tp\gkj X_j \\
\xjb\tp\gkj\xjb &  -\r1\xjb\tp \rj X_j \esmx
\endaligned}
&\cr
\espho
\noalign{\hrule}
\espho
&\mdbox{\frac{{\grs\gro}(2,2l)}{\grs\gro(2) \oplus \grs\gro(2l)}}
&&
\mdbox{
\aligned
\adiu&=\adib \\
\yu&=-\gkti\xdb\iul \\
\xu&=-\gkti\ydb\iul\endaligned}
\separer
\mdbox{i=1,\dots,m}
\separer
\mdbox{
\aligned
\aj&=\ajb \\
Y_j&=-\r1\,\gkj\rj\xjb\iul \endaligned}
\separer
\mdbox{j=2m+1,\dots,n}
&&
\mdbox{2\sm \vtop{
\mdbox{tr(\gkti\vtop{
\mdbox{(\xd \wedge d\yd\tp}\vglue 2pt
\mdbox{\  + d\xdb \wedge d\ydb\tp))}}}
\mdbox{-2\,\r1\sjm \kern -2pt tr(dX_j\wedge
d(\iul\xjb\tp)\rj)}}}
&&
\mdbox{
\aligned
u &= \sm  \vtop{
\mdbox{(\adib\xthb\tp\gkti\ydb^1}
\mdbox{\ \  - \adi\yth\tp\gkti\xd^1)}
\mdbox{\ -\r1 \sjm \aj\xjhb\tp\rj\xjb^1}} \\
\vspace{ 3pt}
v &= \sm \vtop{
\mdbox{(\adib \ythb\tp \gkti \ydb^1}
\mdbox{\ \ -\adi \xth\tp \gkti \xd^1)}
\mdbox{ -\sjm \aj \Xhat_j\tp \rj X_j^1}}
\endaligned}
&&
\mdbox{\ncaliu =}\separer
\mdbox{ \smx \xd\tp\gkti\yd & \kenmi X_{2i}\tp\gkti X_{2i} \\
\yd\tp\gkti\yd & \kenmi\yd\tp\gkti\xd \esmx}\separer
\mdbox{\ncalid =}\separer
\mdbox{ \smx \iul\ydb\tp\gkti\xdb\iul &
\kenmi\iul\ydb\tp\gkti\ydb\iul \\
\iul\xdb\tp\gkti\xdb\iul &
\kenmi\iul\xdb\tp\gkti\ydb\iul \esmx}\separer
\mdbox{\ncalj=}\separer
\mdbox{\smx \r1 X_j\tp\rj\xjb\iul &  X_j\tp\gkj X_j \\
-\iul\xjb\tp\gkj\xjb\iul &  \kern 2pt -\r1\iul\xjb\tp\rj\xjb \esmx}
&
\endboxedalign

\newpage
\centerline{\bf Table $II$c}
\centerline{Darboux Coordinates for CNLS Equations ($BI$)}
\vglue 2pt
\boxedalign
&
\textfrac{Algebra}{+1 eigenspace}
&&\hbox{Reduced coord.}\vglue 2pt \hbox{\& reality cond.}
&&\hbox{Symplectic form}
&&\hbox{$q,\qt,p,\pt,q^1,p^1; u,v,u^1(X,Y)$}
&&\hbox{Moment map $\CalN(\lbd)$ residues$^{1,2}$}
&\cr
\espho
\noalign{\hrule height 0.8pt}
\espho
&\msbox{{{\grs\gro}(2+(2l+1),\cx)}
\over {\grs\gro(2,\cx) \oplus \grs\gro(2l+1,\cx)}}
&&
\mdbox{G_i=(w_i,X_i,Y_i)}\separer
\mdbox{F_i=\gki(w_i,Y_i,X_i)}\separer
\mdbox{w_i\in\cx^{2\kappa_i}}\separer
\mdbox{X_i,Y_i\in\cx^{2\kappa_i\times(l+1)}}\separer
\mdbox{i=1,\dots,n}
&&
\mdbox{\sn tr(\gki(2\,dX_i\wedge dY_i\tp +
dw_i\wedge dw_i\tp))}
&&
\mdbox{\aligned
q &= -\sn\al\Yhat_i\tp\gki X_i^1 \\
\qt &= -\sn\al\Xhat_i\tp\gki X_i^1 \\
p &= \sn\al\Xhat_i\tp\gki Y_i^1 \\
\pt &= \sn\al\Yhat_i\tp\gki Y_i^1 \\
q^1 &= \sn\al w_i\tp\gki X_i^1 \\
p^1 &= -\sn\al w_i\tp\gki Y_i^1
\endaligned}
&&
\mdbox{\ncali=\smx 0 & \kenmi w_i\tp\gki Y_i & \kenmi w_i\tp\gki X_i \\
X_i\tp\gki w_i & \kenmi X_i\tp\gki Y_i & \kenmi X_i\tp\gki X_i \\
Y_i\tp\gki w_i & \kenmi Y_i\tp\gki Y_i & \kenmi Y_i\tp\gki X_i
\esmx}
&\cr
\espho
\noalign{\hrule}
\espho
&\msbox{{{\grs\gro}(2+(2l+1))}
\over {\grs\gro(2) \oplus \grs\gro(2l+1)}}
&&
\mdbox{
\aligned
\adiu&=\adib\\
\yu&=-\gkti\xdb\\
\xu&=-\gkti\ydb\\
\wu&=-\gkti\wzb\endaligned}\separer
\mdbox{i=1,\dots,m}\separer
\separer
\mdbox{
\aligned
\aj&=\ajb\\
Y_j&=-\r1\gkj\rj\xjb\\
w_j&=-\r1\gkj\rj\wjb\endaligned}\separer
\mdbox{j=2m+1,\dots,n}
&&
\mdbox{\sm tr(\gkti\vtop{
\mdbox{(2\,d\xd\wedge d\yd\tp +d\wz\wedge d\wz\tp}
\vglue 2pt
\mdbox{+2\,d\xdb\wedge d\ydb\tp + d\wzb\wedge d\wzb\tp))}}}
\separer
\mdbox{+\r1 \sjm tr\vtop{
\mdbox{(2\,d\xjb\wedge dX_j\tp \rj}
\mdbox{+d\wbar_j\wedge dw_j\tp)}}}
&&
\mdbox{\aligned
u &= -\sm \vtop{\mdbox{ ( \adi\yth\tp\gkti\xd^1}
\mdbox{\quad + \adib\xthb\tp\gkti\ydb^1)}
\mdbox{+\r1\sjm\aj\xjhb\tp\rj X_j^1}} \\
v &=-\sm \vtop{
\mdbox{(\adi \xth\tp \gkti \xd^1}
\mdbox{\ \ +\adib \ythb\tp \gkti \ydb^1)}
\mdbox{ +\sjm \aj \Xhat_j\tp \rj X_j^1}} \\
u^1 \kern -3pt &= \vtop{
 \mdbox{\sm \vtop{
  \mdbox{(\adib\wzb\tp\gkti\ydb^1}
  \mdbox{\ \ \ + \adi\wz\tp\gkti\xd^1)}}}
 \mdbox{+\sjm\aj w_j\tp\gkj X_j^1 }}
\endaligned}
&&
\mdbox{\ncaliu=}\separer
\mdbox{\smx 0 & \kenmi\wz\tp\gkti\yd & \kenmi\wz\tp\gkti\xd \\
\kenmi\xd\tp\gkti\wz & \kenmi\xd\tp\gkti\yd & \kenmi\xd\tp\gkti\xd \\
\kenmi\yd\tp\gkti\wz & \kenmi\yd\tp\gkti\yd & \kenmi\yd\tp\gkti\xd
\esmx}\separer\separer
\mdbox{\ncalid=}\separer
\mdbox{\smx 0   & \kenmi\wzb\tp\gkti\xdb & \kenmi\wzb\tp\gkti\ydb \\
\kenmi\ydb\tp\gkti\wzb & \kenmi\ydb\tp\gkti\xdb & \kenmi\ydb\tp\gkti\ydb \\
\kenmi\xdb\tp\gkti\wzb & \kenmi\xdb\tp\gkti\xdb & \kenmi\xdb\tp\gkti\ydb
\esmx}\separer \separer
\mdbox{\ncalj=}\separer
\mdbox{\smx 0 &
\r1 w_j\tp\rj\xjb &  w_j\tp\gkj X_j \\
 X_j\tp\gkj w_j & \r1 X_j\tp\rj\xjb & X_j\tp\gkj X_j \\
-\r1\,\xjb\tp\rj w_j & \xjb\tp\gkj\xjb & -\r1\,\xjb\rj
X_j \esmx
}
&
\endboxedalign

\newpage
\centerline{{\bf Table $II$c} (cont'd)}
\centerline{Darboux Coordinates for CNLS Equations ($BI$)}
\vglue 2pt
\boxedalign
&
\textfrac{Algebra}{+1 eigenspace}
&&\hbox{Reduced coord.}\vglue 2pt \hbox{\&
 reality cond.}
&&\hbox{Symplectic form}
&&\hbox{$u,v,u^1(X,Y)$}
&&\hbox{Moment map $\CalN(\lbd)$ residues$^2$}
&\cr
\espho
\noalign{\hrule height 0.8pt}
\espho
&
\msbox{{ {\grs\gro}(2,2l+1)}
\over {\grs\gro(2) \oplus \grs\gro(2l+1)}}
&&
\mdbox{
\aligned
\adiu &= \adib \\
\yu &= -\gkti\xdb\iul\\
\xu &= -\gkti\ydb\iul\\
\wu &= \gkti\wzb\endaligned}\separer
\mdbox{i=1,\dots,m}\separer
\mdbox{
\aligned
Y_j &= -\r1\gkj\rj\xjb\iul\\
w_j&=\r1\gkj\rj\wjb\endaligned}\separer
\mdbox{j=2m+1,\dots,n}
&&
\mdbox{\sm tr(\gkti\vtop{
\mdbox{(2\,d\xd\wedge d\yd\tp +d\wz\wedge d\wz\tp}
\vglue 2pt
\mdbox{+2\,d\xdb\wedge d\ydb\tp + d\wzb\wedge d\wzb\tp))}}}
\separer
\mdbox{+\r1 \sjm tr\vtop{
\mdbox{(2\,d(\xjb\iul)\wedge dX_j\tp \rj}
\mdbox{+d\wbar_j\wedge dw_j\tp)}}}
&&
\mdbox{
\aligned
u &= \vtop{
 \mdbox{\sm \vtop{
  \mdbox{(\adi\yth\tp\gkti\xd^1}
  \mdbox{+\adib\xthb\tp\gkti\ydb^1)}}}
 \mdbox{ -\r1\sjm \kern -3pt\aj\xjhb\tp\rj\xjb^1}} \\
v &=\sm \vtop{
\mdbox{(\adib \ythb\tp \gkti \ydb^1}
\mdbox{\ \ -\adi \xth\tp \gkti \xd^1)}
\mdbox{ +\sjm \aj \Xhat_j\tp \rj X_j^1}} \\
u^1 \kern -3pt&=  \vtop{
 \mdbox{\sm \vtop{
  \mdbox{(\adib\wzb\gkti\ydb^1}
  \mdbox{+\adi\wz\tp\gkti\xd^1)}}}
 \mdbox{+ \sjm \aj w_j\tp\gkj X_j^1}}
\endaligned}
&&
\mdbox{\ncaliu=}\separer
\mdbox{\smx 0 & \kenmi\wz\tp\gkti\yd & \kenmi\wz\tp\gkti\xd \\
\xd\tp\gkti\wz & \kenmi\xd\tp\gkti\yd & \kenmi\xd\tp\gkti\xd \\
\yd\tp\gkti\wz & \kenmi\yd\tp\gkti\yd & \kenmi\yd\tp\gkti\xd
\esmx}\separer\separer
\mdbox{\ncalid=}\separer
\mdbox{
\smx 0        & \wzb\tp\gkti\xdb\iul & \wzb\tp\gkti\ydb\iul \\
-\iul\ydb\tp\gkti\wzb & \kern 3pt\iul\ydb\tp\gkti\xdb\iul &
\kern 3pt\iul\ydb\tp\gkti\ydb\iul \\
-\iul\xdb\tp\gkti\wzb & \iul\xdb\tp\gkti\xdb\iul &
\iul\xdb\tp\gkti\ydb\iul
\esmx}
\separer
\mdbox{\ncalj=}\separer
\mdbox{\smx 0 & \wjb\tp\gkj\xjb\iul & w_j\tp\gkj X_j \\
X_j\tp\gkj w_j & \r1 X_j\tp\rj\xjb\iul & X_j\tp\gkj X_j \\
\iul\xjb\tp\gkj\wjb & \kern 3pt -\iul\xjb\tp\gkj\xjb\iul &
\kern 3pt -\r1\iul\xjb\tp\rj\xjb \esmx}
&
\endboxedalign

\noindent
Notation:

\noindent
(1) For the complex Hermitian symmetric Lie algebras:
\mdbox{\Cal N(\lbd) = \sum_{i=1}^n \frac{\lbd {\Cal N}_i}
{\ai-\lbd}}.

\noindent
(2) For their real forms:
\mdbox{\CalN(\lbd)=\sm\left(\frac{\lbd\ncaliu}{\adi-\lbd}
+\frac{\lbd\ncalid}{\adib-\lbd}\right)
+\sjm\frac{\lbd\ncalj}{\aj-\lbd}}.

\vglue 3pt
\hbox{\vtop{\hbox{(3) By
$X_i^1$ (resp.$ Y_i^1$) we denote the first column of
$ X_i$ (resp. $ Y_i$), and by $ \Xhat_i$ (resp.
$\Yhat_i$)  the matrix $ X_i$ (resp. $ Y_i$)
without its first column.\hfil}
\mdbox{(4)\ 2\kappa_i=k_i,\,i=1,\dots,n}
\mdbox{(5)\ \gamma_\kappa=\smx 0 & I_\kappa \\ -I_\kappa & 0
\esmx}
\mdbox{(6)\ I_{k,l}=\smx I_k & 0 \\ 0 & -I_l \esmx}
\mdbox{(7)\ \rho_j=I_{\kappa_j,\kappa_j}}
}}

\end